\documentclass[preprint]{aastex}
\usepackage{graphicx}
\usepackage{url}
\usepackage{hhline}
\usepackage{natbib}
\usepackage{graphicx,epsfig}% Include figure files
\usepackage{bm}% bold math
\usepackage{color}
\usepackage{comment}

\newcommand{\vegamagflp}{$23.20$} 
\newcommand{\vegamagfw}{$24.57$}

\newcommand{\fwfwhm}{1.84} %pixels
\newcommand{\flpfwhm}{1.96} %pixels

\newcommand{\pleiadesage}{$125$} 
\newcommand{\pleiadesdistance}{$136.2$}

\newcommand{\nobjfieldtwoau}{22}

\newcommand{\nobjfieldtenau}{17}

\newcommand{\nbinfieldtwoau}{5}
\newcommand{\nbinfieldfourau}{2}
\newcommand{\nbinfieldtenau}{0}

\newcommand{\binfreqfieldtwoau}{8-19} %separations > 2 AU

\newcommand{\binfreqtaurus}{0.0-6.0}

\newcommand{\binfreqcham}{0.0-10.0}

\newcommand{\binfrequppersco}{0.0-4.0}

\newcommand{\binfreqpleiades}{<$$7.0}

\newcommand{\binfreqfieldtenau}{<$$3.0}

\newcommand{\binfreqtauruslower}{0.0-6.0}

\newcommand{\taupleiades}{125}
\newcommand{\tautaurus}{1}
\newcommand{\taucham}{2-3}
\newcommand{\tauuppersco}{11}
\newcommand{\taufield}{0.5-5.0} %Gyr

\newcommand{\nobjpleiades}{15}
\newcommand{\nobjpleiadeswfconly}{11}
\newcommand{\nobjtaurus}{37}
\newcommand{\nobjcham}{22}
\newcommand{\nobjuppersco}{28}

\newcommand{\nbinpleiades}{0}

\newcommand{\nbintaurus}{3}
\newcommand{\nbincham}{1}
\newcommand{\nbinuppersco}{0}

\newcommand{\qrngpleiades}{$\gtrsim0.6$}

\newcommand{\qrngtaurus}{$\gtrsim0.7$}
\newcommand{\qrngcham}{$\gtrsim0.7$}
\newcommand{\qrnguppersco}{$\gtrsim0.8$}
\newcommand{\qrngfield}{$\gtrsim0.6$}

\newcommand{\fwmedsn}{93.5}%S/N
\newcommand{\fwminsn}{61.1}
\newcommand{\flpmedsn}{49.1} 
\newcommand{\flpminsn}{33.0}

\newcommand{\fwnonmodresid}{5}%percent, modified, mean(abs((data-model)/data)) +/-4
\newcommand{\fwmodresid}{0.9}
\newcommand{\flpnonmodresid}{6} 
\newcommand{\flpmodresid}{2.3}

\newcommand{\binfreqjefftwosig}{<$$26}
\newcommand{\binfreqjeff}{<$$11}

\newcommand{\nguess}{150}
  
\newcommand{\pixlim}{$0.2$}

\newcommand{\nbinary}{$4800$}

\newcommand{\mjup}{\ensuremath{\mbox{M}_{\mbox{\tiny Jup}}}}

\def\astrosun {\mbox{$\odot$}}
\newcommand{\Msol}{\ensuremath{\mbox{M}_{\astrosun}}}

\newcommand{\massrange}{$25-40$~$\mjup$}

\shorttitle{L dwarf Binary Frequency of the Pleiades} 
\shortauthors{Garcia et al.}

\begin{document}

\title{On the Binary Frequency of the Lowest Mass Members of the Pleiades with  {\it Hubble Space Telescope} Wide Field Camera 3}

\author{E.V.\ Garcia\altaffilmark{1,2}, 
Trent J. Dupuy\altaffilmark{3},
Katelyn N. Allers\altaffilmark{4},
Michael C. Liu\altaffilmark{5},
Niall R. Deacon\altaffilmark{6}}

\altaffiltext{1}{Pre-doctoral Fellow Lowell Observatory, 1400 West Mars Hill Road, Flagstaff, eugenio.v.garcia@gmail.com} 
\altaffiltext{2}{Department of Physics \& Astronomy, Vanderbilt University, VU Station B 1807, Nashville, TN 37235, USA} 
\altaffiltext{3}{The University of Texas at Austin, Department of Astronomy, 2515 Speedway C1400, Austin, TX 78712, USA}
\altaffiltext{4}{Department of Physics and Astronomy, Bucknell University, Lewisburg, PA 17837}
\altaffiltext{5}{Institute for Astronomy, University of HawaiÔi, 2680 Woodlawn Drive, Honolulu, HI 96822, USA}
\altaffiltext{6}{Centre for Astrophysics Research, University of Hertfordshire, College Lane, Hatfield, AL1 5TL, UK}

\begin{abstract}
We present the results of a {\it Hubble Space Telescope} Wide Field Camera 3 imaging 
survey of $\nobjpleiadeswfconly$ of the lowest mass brown dwarfs in the Pleiades known (\massrange). These 
objects represent the predecessors to T dwarfs in the field. 
Using a semi-empirical binary PSF-fitting technique, we are able to probe to 
$0\farcs03$ (0.75 pixel), better than 2x the WFC3/UVIS diffraction limit. We did not find any companions to our targets. 
From extensive testing of our PSF-fitting method on simulated binaries, we compute detection limits which rule out 
companions to our targets with mass ratios of $q\gtrsim0.7$ and separations $a\gtrsim4$ AU. 
Thus, our survey is the first to attain the high angular resolution needed to resolve brown 
dwarf binaries in the Pleiades at separations that are most common in the field population. We constrain the binary 
frequency over this range of separation and mass ratio of  \massrange~Pleiades 
brown dwarfs to be $\binfreqjeff\%$ for $1\sigma$ ($\binfreqjefftwosig\%$ at $2\sigma$). 
This binary frequency is consistent with both younger and older brown
dwarfs in this mass range. 
\end{abstract}

\section{Introduction \label{sec:intro}}

Hundreds of brown dwarfs have now been identified in the solar neighborhood
through wide-field surveys (e.g. DENIS, 2MASS, SDSS, UKIDSS, Pan-STARRS and WISE) and in nearby star-forming
regions \citep[e.g.,][]{Epchtein97,Delfosse97,Chiu06,Allers06,Bihain06, Reid08,Bihain10,Burningham10,Cushing11,Liu11,Lodieu12,Burningham13}. 
The study of brown dwarf binarity is a fundamental tool for testing theory, given that the 
statistical properties of binaries probe formation scenarios in the very low-mass regime \citep[e.g.,][]{Burgasser07,Bate09,Luhman12,Bate12}. 
For the past decade, {\it HST} and ground-based adaptive optics (AO) have
fueled such studies by searching for binaries among field ($\taufield$ Gyr) brown dwarfs, 
\citep[e.g.,][]{Martin98,Burgasser03,Bouy03,Burgasser06,Liu06} and in young ($1-10$ Myr) star-forming regions
such as Upper Sco \citep{Kraus05,Bouy06b,Biller11,Kraus12}, Taurus \citep[e.g.][]{Kraus06, Konopacky07,Todorov10,Kraus12,Todorov14},
and Chamaeleon I \citep[e.g.][]{Neuhauser02,Luhman04,Laf08,Ahmic07,Luhman07}. 
Multiplicity studies have also been performed in older ($\approx$$400$ Myr) regions such as Coma Ber, 
Praesepe, and the Hyades \citep{Kraus07,Duch13}. 

Previous work has shown that the binary frequency decreases 
and typical mass ratios increase going to lower mass primaries \citep{Burgasser07}. 
One surprising finding is that
these properties apparently differ between young and old binaries, with the binary frequency
enhanced at young ages by a factor of $\approx$2$\times$ \citep[e.g.,][]{Laf08} and with wide
separations ($\approx$$10-1000$ AU) being much more common as compared to field brown dwarf
binaries that are rarely wider than 10 AU \citep[e.g.][]{Burgasser06,Close07}. An unambiguous physical
explanation for this difference is lacking, as even relatively wide binaries in young star-forming
regions \citep{Luhman04,Luhman09} are not expected to incur dynamical interactions of sufficient intensity to
reduce their frequency and truncate their separation distribution.

The Pleiades open cluster serves as an important bridge between the youngest ($1-10$\, Myr) brown dwarfs
and the field population. It has several advantages, such as its well established 
age~of $\approx$$\taupleiades$ Myr \citep{Stauffer98,Barrado04} 
and distance of $\pleiadesdistance\pm1.2$ pc~\citep{Melis14}. 
There are many surveys that have searched for brown dwarf binaries in the Pleiades 
\citep{Martin00,Dobbie02,Jameson02,Nagashima03,Moraux03,Bouy06}.
However, there are only 4 Pleiades brown dwarfs with primary masses $\lesssim$40~$\mjup$ that have 
been searched for companions to date \citep{Moraux03,Bouy06}. 
At such masses, these objects will cool to T dwarfs at ages of the field population.

In this work, we triple the number of low mass Pleiades brown dwarfs searched for companions, 
surveying a sample of 11 previously unobserved L dwarfs in the Pleiades using {\it HST}/WFC3. We computed
detection limits for our sample using a binary fitting technique and Tiny Tim PSF models. We compared our binary frequency to the observed frequencies for brown dwarfs at similar masses 
in Taurus, Chamaeleon I, Upper Scorpius, and the field population.

\section{Observations \label{sec:obs}} 

\subsection{Sample \label{sec:samp}}
We obtained images of 11 Pleiades brown dwarfs using the Hubble Space Telescope ({\it HST}) with the 
UVIS channel of Wide Field Camera 3 (WFC3/UVIS) in January and February of 2012 (GO 12563, PI Dupuy). 
Our sample consists of the faintest ($K\gtrsim16$ mag), latest type ($\gtrsim$M9) members 
of the Pleiades known in early 2011. According to BT-Settl models of \cite{Allard14} tied to the COND evolutionary models of \cite{Baraffe97,Baraffe98,Baraffe03}, the estimated masses of our sample 
are \massrange~based on their $K-$band magnitudes and the age of the Pleiades. When defining our sample, we considered objects bona-fide members of the Pleiades if they had proper motion indicating cluster membership and spectra with 
low surface gravity features or lithium absorption. Our sample is listed in Table~\ref{table:samp}, along 
with 4 targets from previous {\it HST}/ACS and {\it HST}/WFPC2 observations of Pleiades 
brown dwarfs by \cite{Martin03} and \cite{Bouy06} that match our membership criteria. All of our sample have
 proper motions consistent with the Pleiades cluster \citep{Bihain06,Casewell07,Lodieu12}. 
 BRB 17, BRB 21, PLIZ 35, BRB 23 and BRB 29 have spectral types L0-L4.5 from \cite{Bihain10}. 

\subsection{{\it HST}/WFC3 Imaging \label{sec:hstwfc3imaging}}
We obtained 2 exposures each in filters F814W and F850LP
for each target star. One image of brown dwarf BRB 17 was lost due to a pointing error so we had a total of 43 images. 
The target stars are positioned near the center of the full field of view at 
$\approx$$250$ pixels from the bottom of chip 1.  We chose a longer exposure time of 900 s in F814W filter, 
where we are sensitive to tighter brown dwarf binaries because of the smaller PSF. We also obtained 340 s exposures in F850LP to 
confirm the presence of any candidate companions and measure their colors. 
The full width half maximum of the PSF  is $\approx$$\fwfwhm$ pixels in F814W
and $\approx$$\flpfwhm$ pixels in F850LP according to 
the WFC3 data handbook\footnote{\url{http://www.stsci.edu/hst/wfc3/documents/handbooks/currentIHB/}}.

We inspected each image for cosmic rays hits, identified as rays or streaks 
with high counts but not resembling WFC3 point sources. We found 6 of the 43 images had cosmic 
ray hits within 5 pixels of the target star. We use the Laplacian Cosmic Ray Identification 
algorithm {\tt LACOSMIC} \citep{Dokkum01} to remove cosmic rays from a 200$\times$200 pixel area 
on the detector centered on the target star. {\tt LACOSMIC} replaces each pixel with the median of the surrounding 
pixels in an iterative procedure. Visual inspection after the fact confirms that we successfully cleared
all obvious cosmic ray hits except for a single image of brown dwarf BRB 23 in F850LP 
due to a cosmic ray hit through the center of the peak of the target. 
We excluded this image of BRB 23 in the subsequent data analysis, therefore leaving us with 42 images total for the rest of
our analysis. 

We computed aperture photometry of our targets from the pipeline calibrated, geometrically-corrected, 
dither-combined (drz) images.  We calculated our aperture photometry using the {\tt APER} task from the 
IDL Astronomy User's Library\footnote{\url{http://idlastro.gsfc.nasa.gov/homepage.html}} for an aperture radius of $0\farcs4$ and a sky annulus of $0\farcs4-0\farcs8$.  We converted the flux in our aperture to a Vega magnitude 
using zeropoints of~\vegamagfw~mag for the F814W filter and~\vegamagflp~mag for the F850LP filter provided in the {\it HST}/WFC3 webpages\footnote{\url{http://www.stsci.edu/hst/wfc3/phot_zp_lbn}}.  
To determine our photometric uncertainties, we first constructed an error image for each image, 
accounting for read noise and poisson noise.  Using a Monte Carlo approach, we determined our 
photometric errors from $10^{4}$ iterations of the {\tt APER} task after adding random Gaussian 
noise to the image in each iteration. The resulting F814W and F850LP photometry for our targets is listed in Table~\ref{table:phot}.

\section{Image Analysis \label{sec:imanalysis}}

\subsection{Point Spread Function (PSF) Model of WFC3/UVIS \label{sec:psfmod}}

In order to search for close companions to our targets, we began by fitting a model Tiny Tim \citep{Krist11} 
point spread function to our imaging data. To create the most 
accurate model we specified the exact coordinates of our target and used an 
input spectrum of 2MASS J00361617+1821104 \citep[][L3.5]{Reid00}. 
We set the defocus parameter in Tiny Tim to the model defocus provided on the Space Telescope 
Science Institute webpage\footnote{\url{http://www.stsci.edu/hst/observatory/focus}} for each image of each target. The 
model defocus is computed to account for breathing, according to the telescope temperature data. 

We sampled the Tiny Tim PSF at 5$\times$ 
the pixel scale ($0\farcs04$ pixel$^{-1}$) of WFC3/UVIS1. 
To simulate sub-pixel shifts of our targets we bilinearly interpolated to an arbitrary fractional pixel
and then binned down to pixel scale of WFC3. We used the Nelder-Mead 
downhill simplex method from \cite{NR}, which is the {\tt AMOEBA} algorithm in IDL,  to minimize the $\chi^{2}$,  
varying the $(x,y)$ position and flux normalization until finding the 
best fit. We computed $\chi^{2}$ as ((image-model)/noise)$^{2}$, where 
``noise'' is the noise image provided by the WFC3 reduction pipeline. 
We ran the {\tt AMOEBA} algorithm twice, starting the second 
run at the end point in parameter space of the first run, as recommended by \cite{NR}. We fit a $\pm10$ pixel 
cutout region centered on the target star. 

\begin{figure}[ht]
\begin{center}
\includegraphics[width=\textwidth]{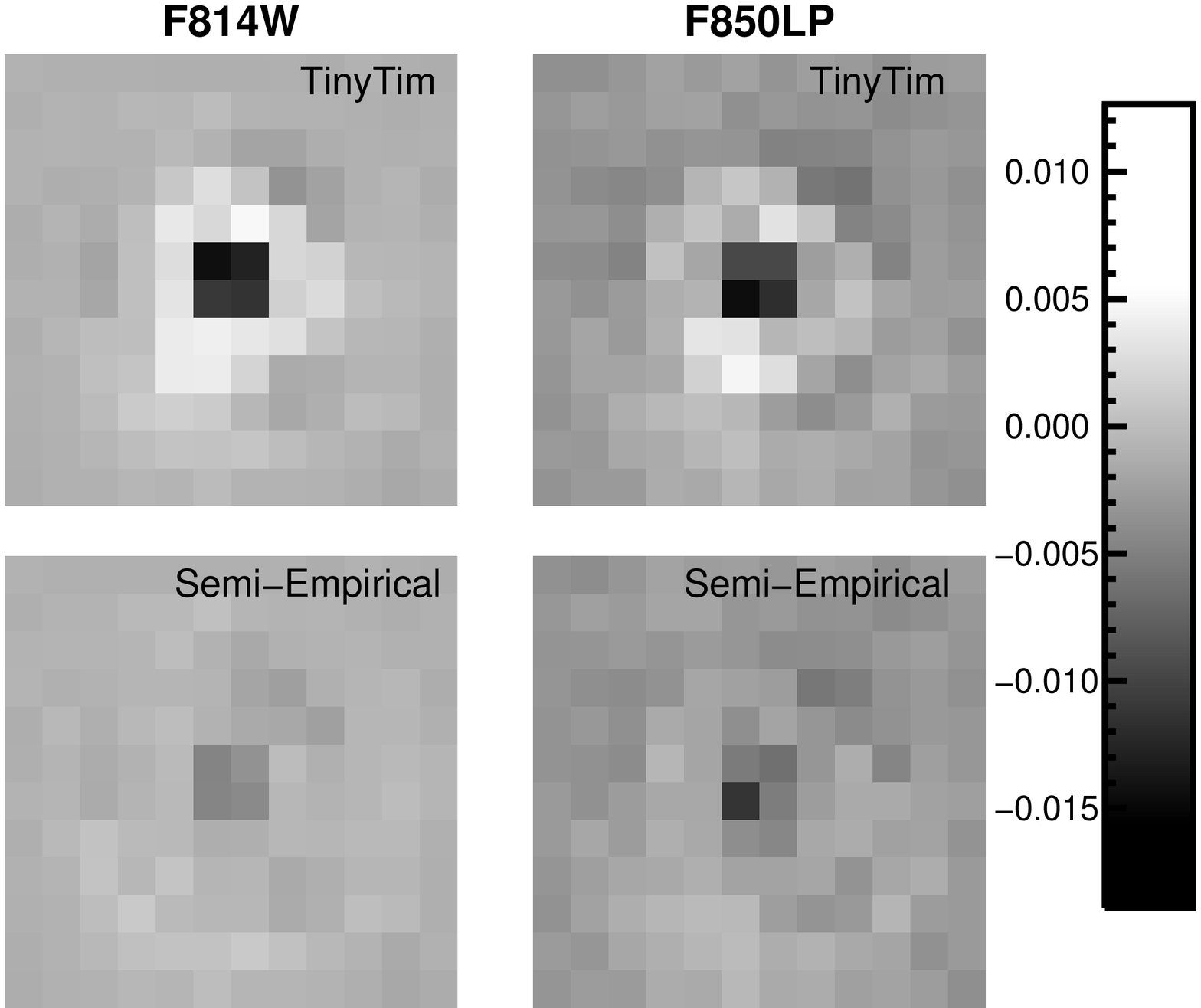}%{residcompare.ps} %/home/garciaev/PleidesBinaries/tinytim/SinglePSFS_080612/plotresid.pro
\end{center}
\caption{\label{fig:resid} 
The average residuals of all 
WFC3 images after fitting the original Tiny Tim model (top) and fitting 
our semi-empirical PSF model (bottom).  For viewing purposes, we display the 
average residuals as the normalized ``(image-model)'' in each filter. 
When using our original Tiny Tim model, the average residuals results in concentrated residual 
flux at instrumental position angles 
of $30-50^{\circ}$ and $150-170^{\circ}$ for both the F814W and F850LP filters.  
This would bias our binary fitting technique to preferentially 
recover companions with these position angles. Therefore we computed 
a semi-empirical model PSF using the original Tiny Tim model as a starting point as detailed in 
\S\ref{sec:psfmod}. The resulting average residuals are improved by a factor of 4-5$\times$ from $\approx\fwnonmodresid\%$ and 
$\approx\flpnonmodresid\%$ to $\approx\fwmodresid\%$ and 
$\approx\flpmodresid\%$ in F814W and F850LP respectively. The residuals are also smoother, no longer containing 
concentrations at position angles of $30-50^{\circ}$ and $150-170^{\circ}$.
}
\end{figure}

We found average residuals after subtracting the best-fit Tiny Tim model of $\fwnonmodresid\%$ and 
$\flpnonmodresid\%$ for F814W and F850LP images, respectively.  
We computed residuals of our fits as the average fractional
offset between the image and the model. The majority of the residual flux using the Tiny Tim model was
at instrumental position angles of $30-50^{\circ}$ and $150-170^{\circ}$ in both the 
F814W and F850LP filters (Figure~\ref{fig:resid}). If we searched
for faint companions using the TinyTim PSF model and our binary fitting technique detailed below, we found that this 
systematic residual flux led to spurious detections of companions at these position angles. 

%normalized residuals
Therefore, we instead computed a single optimal semi-empirical 
PSF model that minimized the residuals across all images by modifying the Tiny Tim model. 
We iteratively solved for a 5$\times$ 
over-sampled additional component image to be added to the Tiny Tim model. 
The best guess of this additional component at 
each pixel was computed as the median across all normalized images of the data minus the previous iteration's PSF model. 
We computed a semi-empirical PSF model as the 
Tiny Tim model at the mean position of our targets with this additional component added in.  
%why weren't all the science coord at the same X, Y? 

Using our semi-empirical PSF model, the final residuals of our fits were improved by 5$\times$ to $\approx$$\fwmodresid\%$ and 
$\approx$$\flpmodresid\%$ for F814W and F850LP, respectively (Figure 
\ref{fig:resid}). Most importantly, we no longer see the concentrated residual flux at 
position angles of $150-170^{\circ}$. We use our semi-empirical PSF model in all subsequent analysis.

The method of fitting binaries is the same as described above, but instead of using a single model
we use two co-added models. As before, the {\tt AMOEBA} algorithm minimizes the $\chi^2$ between 
the image and co-added semi-empirical PSFs. We varied six binary parameters: the primary's position on detector, 
the flux normalization between the primary star and the PSF model primary, the binary separation, 
the position angle, and flux ratio between the primary and secondary. 

\subsection{Quantifying False Positives \label{sec:fpcc}}

If we run our 
binary fitting code on a image of a single star, 
we recover binary parameters of false positive companions. By definition, 
these detections reveal the distribution in separation and flux ratio of the 
false positives we would find while searching for companions in our imaging data. 
To characterize the false positives for our WFC3 data, we  
fit images of our target stars using our binary fitting technique from \S\ref{sec:psfmod}. 
We scale all images to either
the median or minimum S/N of our sample
by adding in gaussian noise (Table~\ref{table:sim}). This allows 
us to put our sample on a common scale for our simulations. 
For each target star, we start with~\nguess~random initial 
guesses, uniformly distributed in ($x,y$) from $0.1-5$ pixels, 
and flux ratios from $0-5$ mag. 

\begin{figure}[ht]
\begin{center}
\includegraphics[angle=90,width=1.5\textwidth]{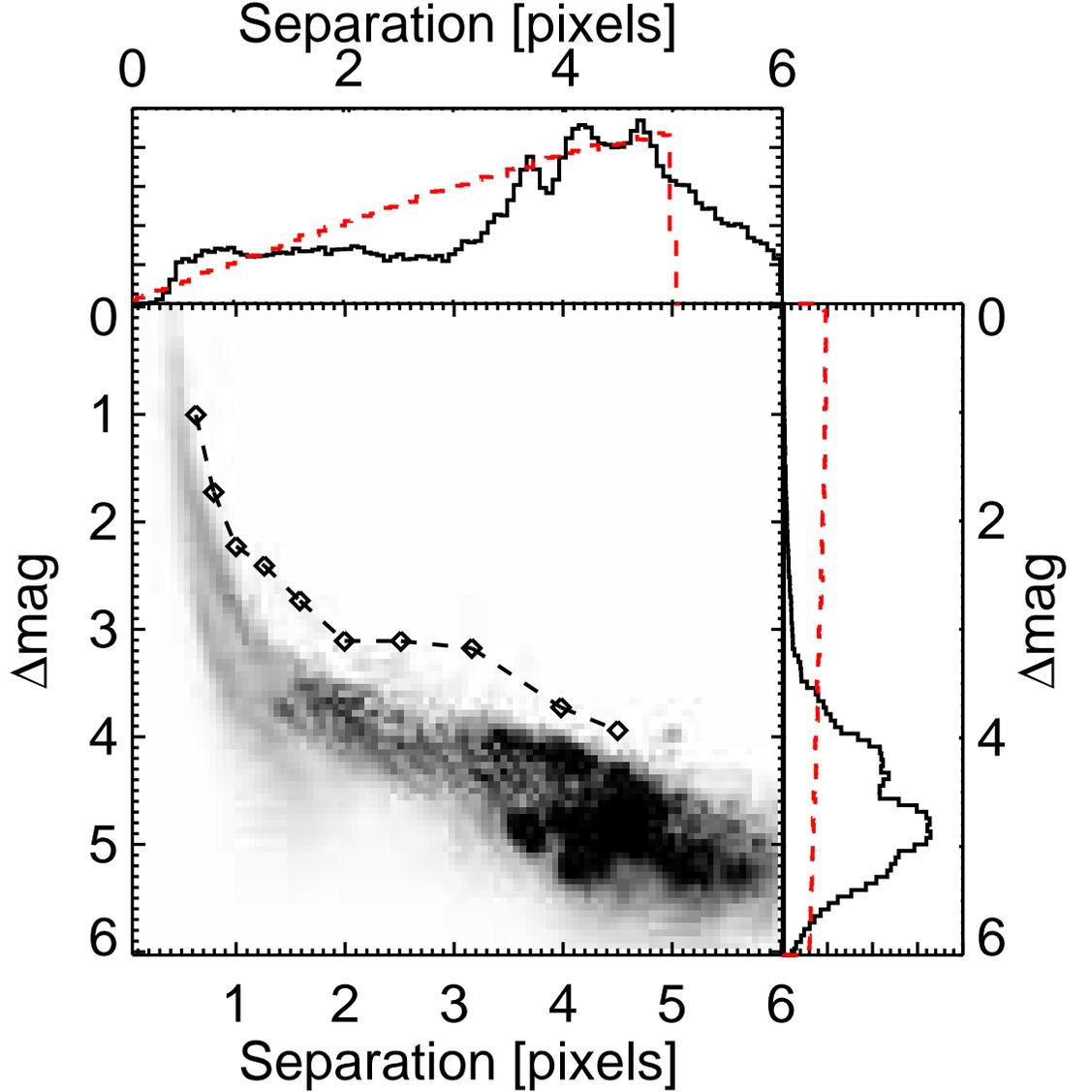}%{F814Wmed_fp.ps} %hstconcurve.pro
\end{center}
\caption{\label{fig:f814wmedhistfp} 
Number density of binary parameters returned when fitting images of single stars with a binary PSF model, i.e., 
false positive detections, for the median S/N F814W case. Overall, the most common false 
positives have wide with separations of $>$3 pixels, and faint flux ratios of $>$4 mag companions, 
but at separations of $<$1 pixels ($<$$0\farcs04$), 
the majority of false positives range with flux ratios of $0-3$ mag. The dotted red histograms
are the initial guesses for the false positives uniformly distributed in log separation 
and flux ratio. The 1\% false positive curve (\S\ref{sec:falspos}) is over plotted (diamonds). 
}
\end{figure}

We show the resulting distribution of separations vs flux ratios  
of recovered false positives in Figure \ref{fig:f814wmedhistfp}. The brightness of false positives increases with 
decreasing separation. At the tightest separations ($<$$0\farcs02$, $<$$0.6$ pixel), 
we find that near unity flux ratio false positives are the most common. 
At wider separations ($>$1.5 pixels, $>$$0\farcs06$), 
we find that almost all false positives are found with large flux ratios of $3-5$ mag.  
This is expected, as the binary fitting code is required to return a position and flux normalization
for a secondary even if one doesn't exist. In other words, the single WFC3 PSF can be fit with
a model of a high flux PSF and a very low flux PSF added 
in to fit any small leftover residuals. 

\subsection{Artificial Binary Simulations \label{sec:sim}}

In order to compute detection limits for our survey, we generated artificial binaries at 
random separations of $0.3-5$ pixels ($0\farcs018-0\farcs2$), position angles of
$0-360^{\circ}$, and flux ratios of $0-5$ mag. We created these artificial 
binaries by shifting, scaling and co-adding randomly selected pairs of actual images together.  
Given that the marginally sampled WFC3 PSF (FWHM~$\lesssim$2 pixels) hinders the accuracy of linear interpolation at sub-pixel shifts, 
we shift the secondary star relative to the primary star in integer pixel steps.  We scaled the image of every primary to 
a common S/N by adding noise, thus degrading the image to lower S/N. We scaled the secondary to a S/N appropriate for the randomly chosen flux 
ratio of the artificial binary. 

Given the integer pixel shifts, there are fixed separations and position angles allowed by the possible image pairings. 
These integer pixel shifts can result in non-integer artificial binary separations because the sub-pixel position for each image varies. 
Out of all possible pairings we selected a subset of~\nbinary~artificial binaries that are distributed uniformly in log separation, 
flux ratio, and position angle. We ran two sets of simulations for each filter, scaling primaries alternatively to the median S/N 
and the minimum S/N of our images (Table~\ref{table:sim}). Only half the images were used for the median S/N simulations, 
given that we only scaled images down in S/N, never up.  

We then blindly fitted for the binary 
parameters of our artificial binaries using a double PSF model as described in \S\ref{sec:psfmod},
using \nguess~random initial guesses. The best-fit values for 
each parameter are calculated as the mean of the resulting~\nguess~runs of our binary 
fitting code parameters where runs with outlier $\chi^{2}$ were excluded from the average. 

\subsection{Deriving False Positive Curves \label{sec:falspos}}

%====================
%FALSE POSITIVE CURVE
%=====================

The binary parameters recovered in our artificial binary simulations contain a mix of both detections and false positives. To assess the likelihood of a given binary fit being a detection, we compared our 
distribution of false positives from \S\ref{sec:fpcc} and our fits to artificial binaries from \S\ref{sec:sim} to 
measure our false positive curve, i.e the largest flux ratio before 
the recovered secondary star becomes indistinguishable from a false positive at a given separation. 

\begin{figure}[ht]
\begin{center}
\includegraphics[angle=90,width=\textwidth]{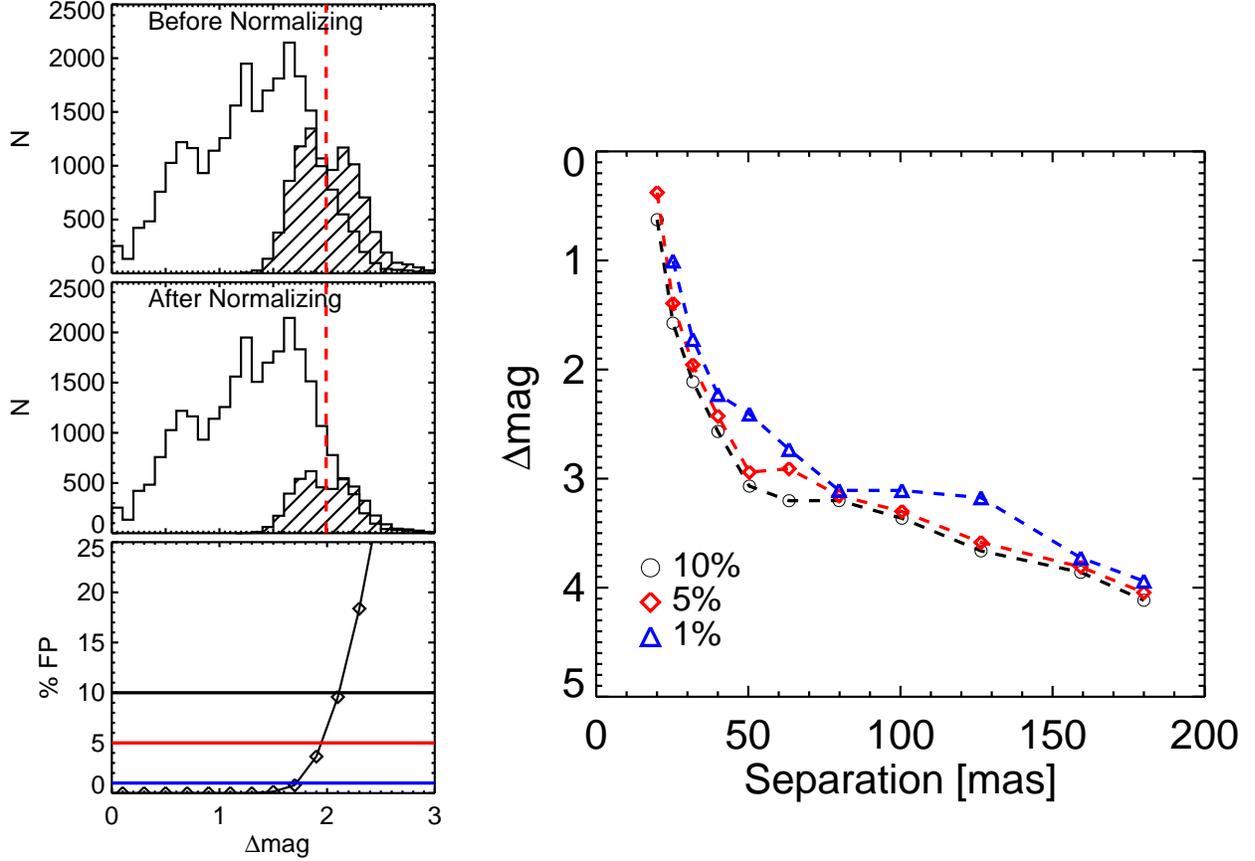}%{process_cc.ps} %/plotpapers/process_cc.pro
\end{center}
\caption{\label{fig:processfp} 
Illustration of our calculation of the false positive curve using the case of artificial binaries at separations $0.79-1.0$ pixels ($\approx0\farcs03-0\farcs04$) as an example.  {\bf TOP LEFT:} The white histogram is the distribution of the recovered flux ratios $\Delta\rm m$ for artificial binaries at separations of $0.79-1.0$ pixels. The histogram with slashes are false positives recovered by using our binary fitting technique on single star images. The vertical red dashed line is the median false positive flux ratio. {\bf MIDDLE LEFT:} We normalize the histogram of false positive flux ratios (slashes) to the white histogram of recovered artificial binary flux ratios by conservatively assuming that any artificial binaries 
with recovered flux ratios larger than the median false positive flux ratio (vertical dashed red line) are most likely false positives themselves. {\bf BOTTOM LEFT:} 1\% (black solid line), 5\% (red) and 10\% (blue) false positive fractions as a function of flux ratio. {\bf RIGHT:} The false positive curve is constructed by repeating the process for all separation bins. The stars denote the $\Delta\rm m$ corresponding to $1\%, 5\%$ and $10\%$ false positive fraction at separations of $0.79-1.0$ pixels shown at the bottom left.}
\end{figure}

We considered the artificial binaries and false positives 
in a given separation and flux ratio range, using 0.1 dex pixel bin widths and $0.3$ mag flux ratio bin widths, respectively. 
In each separation bin we normalized the histogram of false positive flux ratios to the histogram of recovered artificial binary flux ratios by conservatively assuming that any artificial binaries 
with recovered flux ratios larger than the median false positive flux ratio $\Delta\rm m_{\rm crit}$
were most likely false positives themselves. We computed this normalization factor as  $\frac{n}{0.5 n_{\rm fp}}$, where 
$n_{\rm fp}$ is the total number of false positives and $n$ is the number of artificial binaries with flux ratios $>$$\Delta\rm m_{\rm crit}$. 
After normalization, we computed the false positive fraction as a function of flux ratio as $1 - \frac{n_{\rm fp}}{n}$. We repeat 
the procedure above for each separation bin. This procedure is depicted 
in Figure~\ref{fig:processfp} for the $0.79-1.0$ pixel separation bin. 

\begin{figure}[ht]
\begin{center}
\includegraphics[angle=90,width=\textwidth]{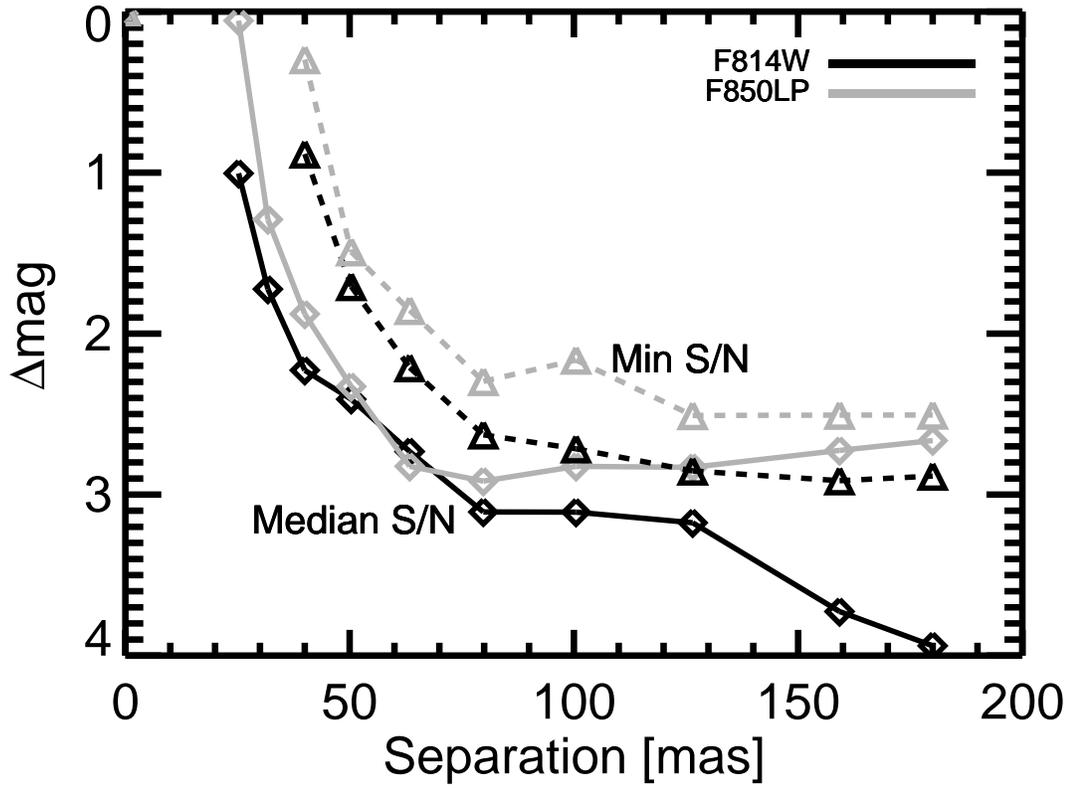}%{falspos_curve.ps} %plots_for_paper.pro
\end{center}
\caption{\label{fig:falpos} 
False positive curves computed at the 
minimum (triangles) and median (squares) 
signal-to-noise of our WFC3 images of Pleiades brown dwarfs (\S\ref{sec:falspos}) for 
the F814W (black) and F850LP (grey) filters. As expected the minimum S/N false 
positive curves have brighter false positives than the 
median S/N curves in a given filter. 
}
\end{figure}

With the procedure detailed above, 
we computed false positive curves at the median and minimum S/N of our images for 
the F814W and F850LP filter as shown in Figure \ref{fig:falpos}. Each of our false positive curves are representative 
of a {\it single}, S/N given that we scale our all our images to 
a common S/N for each set of simulations. 

\subsection{Deriving Contrast Curves \label{sec:cc}} 

We computed contrast curves that correspond to the largest flux ratio companion that our binary PSF
fitting technique can recover accurately at a given separation. 
A binary is considered ``recovered'' 
if the best fit parameters are within~\pixlim~pixels and  $1$ mag 
of the input $(x,y)$ positions and flux ratio, respectively. We binned our simulated binaries by 
separation and flux ratio with bin widths of 0.1 dex pixels and $0.3$ mag, respectively. 
In each bin, we computed the completeness fraction as 
the number of artificial binaries that are recovered divided by the total number of artificial binaries in the bin. 
We define our contrast curves as the flux ratio 
bin at a given separation where the completeness fraction is $90\%$ 
determined by the interpolation of the binned results. We computed contrast curves at
the median and minimum S/N of our targets (Table~\ref{table:sim}) 
for the both F814W and F850LP filters. 

\begin{figure}[ht]
\begin{center}
\includegraphics[angle=90,width=\textwidth]{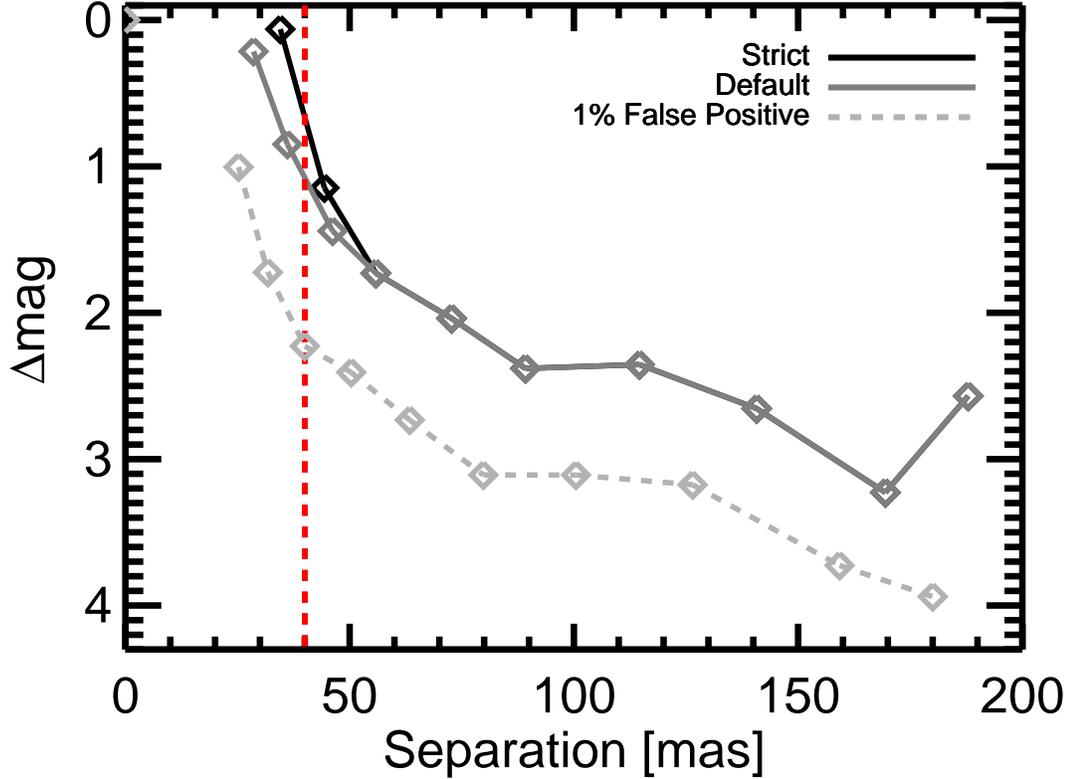}%{concurve.ps} %plots_for_paper.pro
\end{center}
\caption{\label{fig:cc}
Contrast curves for the median F814W S/N case constructed 
using our default and a stricter companion recovery criteria (see \S\ref{sec:cc}). 
In our default criteria, we required companions be recovered to within 
$1$ mag of the input binary parameters (dark grey solid line). 
We can recover companions with flux ratios $<$1 mag and separations 
$>$$0\farcs04$. We also tested a stricter criteria, 
and required recovered companions to be within
within $0.3$ mag of the input flux ratio (black solid line). 
The contrast curves are identical for separations $>$$0\farcs055$.
With the stricter recovery criteria, companions with separations 
$<$$0\farcs04$ and flux ratios $< 0.5$ mag were detectable. 
Both contrast curves required that recovered artificial 
binaries be within~\pixlim~pixels of input ($x,y$) position. We adopt 
our default criteria given that most brown dwarf binaries are found to have 
near unity flux ratios. The 1\% false positive curve is shown for comparison (light grey dotted line). 
The contrast curve drops at 200 mas due to difficulty in 
fitting artificial binaries at the edge of our cut-out region of $\pm10$ pixels. 
 }
\end{figure}

Figure~\ref{fig:cc} shows our resulting contrast curves. We are able to recover tight ($<$$0\farcs04$, $<$1 pixel) binaries 
with flux ratios $\lesssim$1 mag. At wider separations we recover binaries $3-5$ magnitudes fainter. 
We also constructed a contrast curve with a stricter recovery
requirement to be within $0.3$ mag of the input. This leads to a contrast curve that 
reaches in to binary separations of $0\farcs035$ (0.9 pixels) 
and is identical to our default recovery requirements outside $0\farcs055$ (1.4 pixels). A flux ratio of $\lesssim$1 mag for 
our targets corresponds to a mass ratio $q\gtrsim7$ which allows us to rule out the possibility 
of Pleiades brown dwarf binaries similar to field brown dwarf binaries, since the 
latter mostly have $q\approx1$ \citep[see review by][]{Burgasser07}. This
means that a stricter flux ratio requirement of $<$$0.3$ mag for 
constructing our contrast curves is unnecessary. Thus, our PSF fitting technique is able to 
recover artificial binaries as tight as $0\farcs03$, well inside the diffraction 
limit ($\approx\frac{1}{3}\lambda/D$)

Given that each target in our sample has a different S/N, we interpolated over 
the measured median and minimum S/N curves to compute 
a contrast curve for each target. We conservatively fixed the contrast curve 
for our targets with S/N higher than the median S/N to the median S/N contrast curve. 
Our detection limits in F814W and F850LP mag for each target are shown in Table~\ref{table:detectlim}.  
These detection limits have lower contrast and are more conservative 
than the false positive curves, as expected. 
Finally, we convert our contrast curves from F814W and F850LP magnitudes to masses using BT-Settl models 
\cite{Allard14} tied to the COND evolution models of \cite{Baraffe03}. We assumed an age of 
\pleiadesage~Myr \citep{Barrado04} and distance to the Pleiades of \pleiadesdistance~
pc \citep{Melis14}. Figure 6 shows the
90\% completeness contrast curve for each target as a function 
of mass ratio ($q$) and projected separation ($a$) in AU. We use only the F814W contrast curve for our constraint 
on the binary frequency due to higher S/N, larger contrast, and closer limiting separation than our F850LP 
contrast curve. 

\subsection{Completeness Maps\label{sec:compmap}}

Similar to how we derive contrast curves in \S\ref{sec:cc}, we derive a median 
and a minimum S/N completeness map for the F814W and F850LP filters. Each completeness 
map represents the probability that a companion with a given separation 
and flux ratio would have been detected (Figure~\ref{fig:compmaps}). The 
procedure for deriving completeness maps is exactly as deriving a contrast curve in
\S\ref{sec:cc} except that we compute the completeness fraction at every separation 
and flux ratio bin. We computed a completeness map for each target similar to \S\ref{sec:cc}, by 
interpolating over the median and minimum S/N completeness maps. We 
conservatively fixed the completeness maps for our targets with S/N 
higher than the median S/N to the median S/N completeness map. Our completeness 
maps for several targets are shown in Figure~\ref{fig:compmaps}. 

\section{Results \label{sec:results}}

\subsection{L Dwarf Binary Frequency of the Pleiades \label{sec:binfrac}} 

\begin{figure}[ht]
\begin{center}
\includegraphics[angle=90,width=\textwidth]{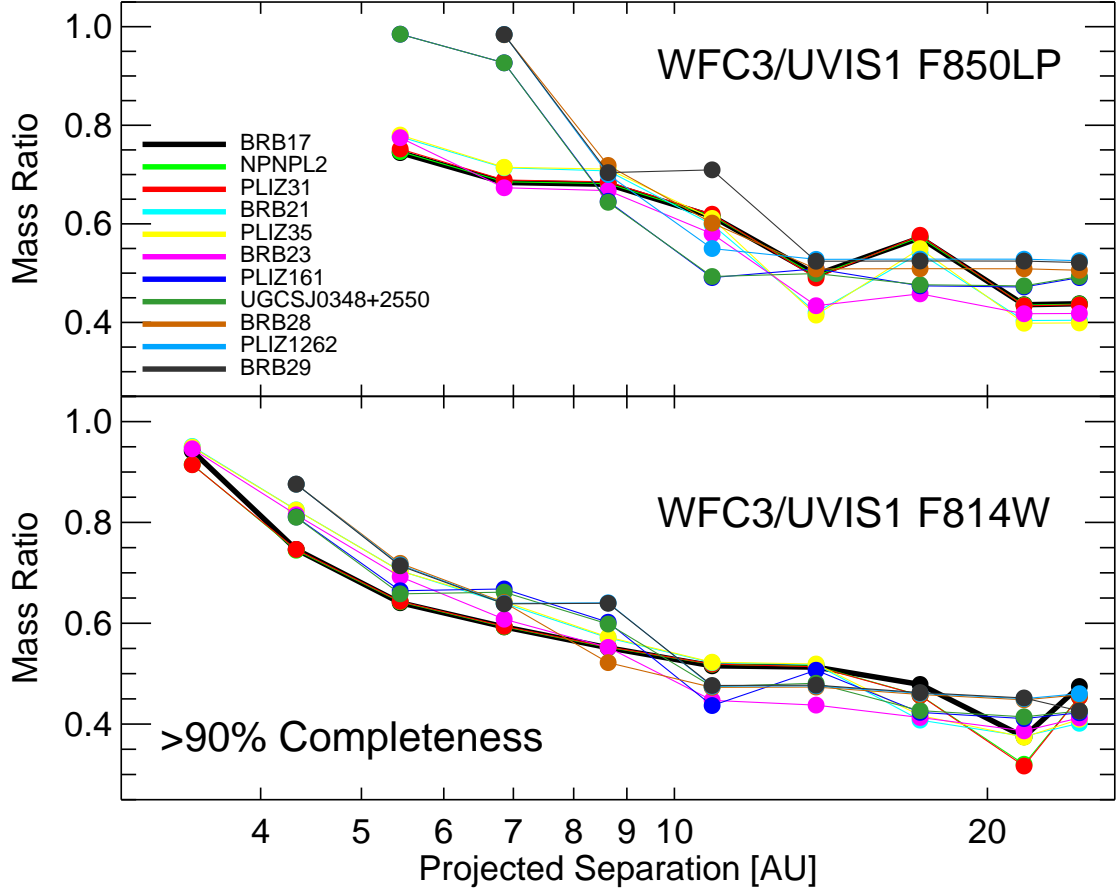}%{allcc.ps} %makecc_combine_edit.pro, don't use makecc_vs_mratplot.pro! 
\end{center}
\caption{\label{fig:allcc}
90\% completeness contrast curves for our F814W and F850LP observations of 11 young L dwarfs ($\lesssim$40$~\mjup$) in the Pleiades. 
Our contrast curves rule out the majority of expected brown dwarf binaries, 
given that most binaries in the field have mass ratios $\gtrsim$$0.6$ and 
separations $<$25 AU \citep{Burgasser07}. We convert our detection limit flux ratios in WFC3 bandpasses
to mass ratios using the distance to the Pleiades \citep[\pleiadesdistance~pc,][]{Melis14} 
and evolution models from~\cite{Baraffe03} tied to BT-Settl models \citep{Allard14}. 
}
\end{figure}

\begin{figure}[ht]
\begin{center}
\includegraphics[angle=90,width=\textwidth]{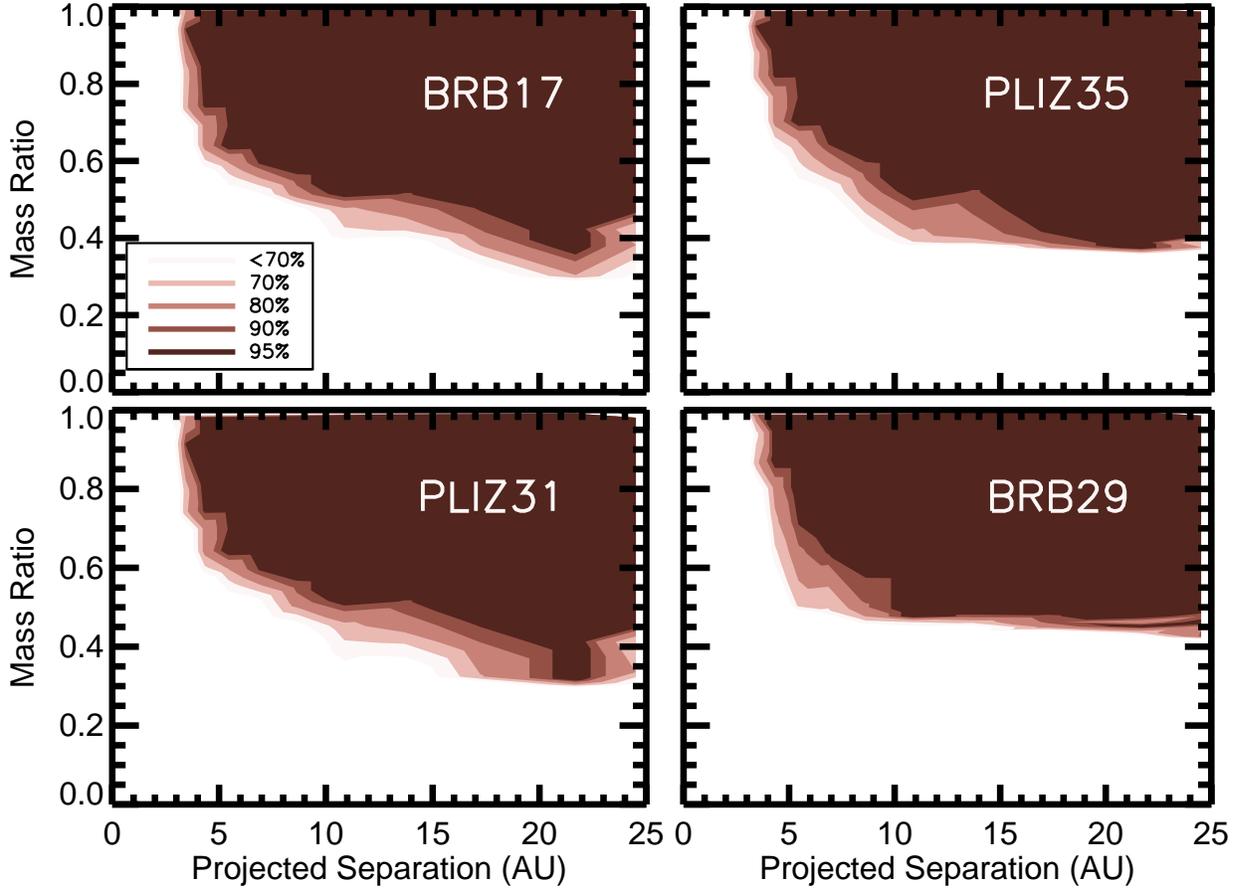}%{compmaps.ps} %plotcompmaps_forpaper 
\end{center}
\caption{\label{fig:compmaps}
Example completeness maps for several our 11 young L dwarfs ($\lesssim$40$~\mjup$) in the Pleiades. 
At each point, the completeness map represents the percentage of binaries that would have been recovered 
given our observations. We convert our detection limit flux ratios in WFC3 bandpasses
to mass ratios using the distance to the Pleiades \citep[\pleiadesdistance~pc,][]{Melis14}, 
and evolution models from~\cite{Baraffe03} tied to BT-Settl models \citep{Allard14}. 
}
\end{figure}

We found no companions in surveying 11 brown dwarf members of the Pleiades with $K\gtrsim16.0$ mag. 
Our F814W contrast curves demonstrate that we could have detected 
companions with mass ratios of $q\gtrsim0.5$ at separations $a\gtrsim10$ AU and $q\gtrsim0.8$ 
at $a\gtrsim4$ AU (Figure~\ref{fig:allcc}). Most known very low mass  binaries are sharply peaked towards 
mass ratios $q\approx1$ \citep{Burgasser06,Liu10}. Furthermore, our detection limits probe 
down to separations $a\approx4$ AU, 
near the peak of the observed binary distribution \citep{Burgasser06}. Thus, our detection limits 
are sensitive to the majority of binaries expected from the observed field population of T dwarfs
\citep{Burgasser03,Burgasser06,Gelino11,Liu12,Radigan13}.  

We estimated the binary frequency for the Pleiades by comparing our completeness maps (\S\ref{sec:compmap}) to 
various random simulated populations of binaries.  Each population of binaries had an
adopted eccentricity, mass ratio and separation distribution, with semi-major axes of 
$<$$25$ AU in accordance with observations of T dwarf binaries in 
the field. We adopted a uniform eccentricity 
distribution of $0-0.9$ in accordance with observations \citep{Dupuy11}. For our mass ratio distribution, 
we used the observed power law of $P(q)\propto q^{4.9}$ \citep{Liu10}. For 
our separation distribution, we used the log normal distribution 
from \cite{Allen07}. We assumed uniform prior distributions of longitude of
ascending node, mean anomaly, and argument of periapsis, and an $a\sin{i}$
distribution for inclination. We projected each binary on sky from the population
with $10^{5}$ randomly chosen 
orbits. We compared each of these $10^{5}$ orbits 
to each completeness map of each target. The probability for detecting a
binary was given by our completeness fraction at the separation 
and mass ratio of the binary from the completeness maps (Figure~\ref{fig:compmaps}). 
We averaged over all probabilities and computed a single average probability (``detectability'')
to recover a companion for each target star (Table~\ref{table:compmapprob}). 
Similar to \cite{Aberasturi14}, we then summed over these average probabilities, and
found that if all our targets had companions we should
have detected $7.6$ binaries for the log normal distribution of semi-major axes. We also used a 
linear (flat) semi-major axis distribution to be consistent with \cite{Aberasturi14}, finding virtually no difference 
in the total number of binaries we should have detected ($8.1$). 
The lack of detections implies a binary frequency upper limit of $\binfreqjeff\%$~
for $1\sigma$ ($\binfreqjefftwosig\%$ at $2\sigma$) using the recommended 
Jeffrey's distribution for small $n$ \citep{Brown01}. 
\cite{Aberasturi14} computed a binary frequency for $\gtrsim$T5 
primaries in the solar neighborhood of $<$16$\%$-$<$25$\%$ using the 
Clopper-Pearson interval at $95\%$ confidence using the same log normal and uniform separation 
distributions. This is comparable 
to our own binary frequency upper limit of $\binfreqjefftwosig\%$ at $2\sigma$ 
($\approx95\%$ confidence).

\subsection{Binary Frequency vs Age for Wide ($>$10 AU) Companions \label{sec:binvsage}}

According to the evolution models of \cite{Baraffe03}, our sample of Pleiades 
L dwarfs are expected to evolve to $T_{\rm eff} = 700-1300$\,K (i.e., T0-T8 spectral types) at ages of 
$0.5-5.0$ Gyr.  At younger ages of $1-10$\,Myr, 
our sample would have had temperatures of $2300-2750$\,K (i.e. $\rm M7-\rm M9$). 
Thus, we compared our binary frequency constraint to AO and {\it HST} observations 
of $\gtrsim$M7 objects in Taurus \citep{Todorov14,Kraus12,Kraus06,Konopacky07,Todorov10},
Chamaeleon I \citep{Luhman04,Laf08,Ahmic07,Luhman07,Neuhauser02}, Upper Sco \citep{Biller11,Kraus12}
and the field \citep{Burgasser06}. Taurus, Chamaeleon I and Upper Sco are regions with objects 
all at the same distance, thus aiding the comparison. 

\begin{figure}[ht]
\begin{center}
\includegraphics[angle=90,width=\textwidth]{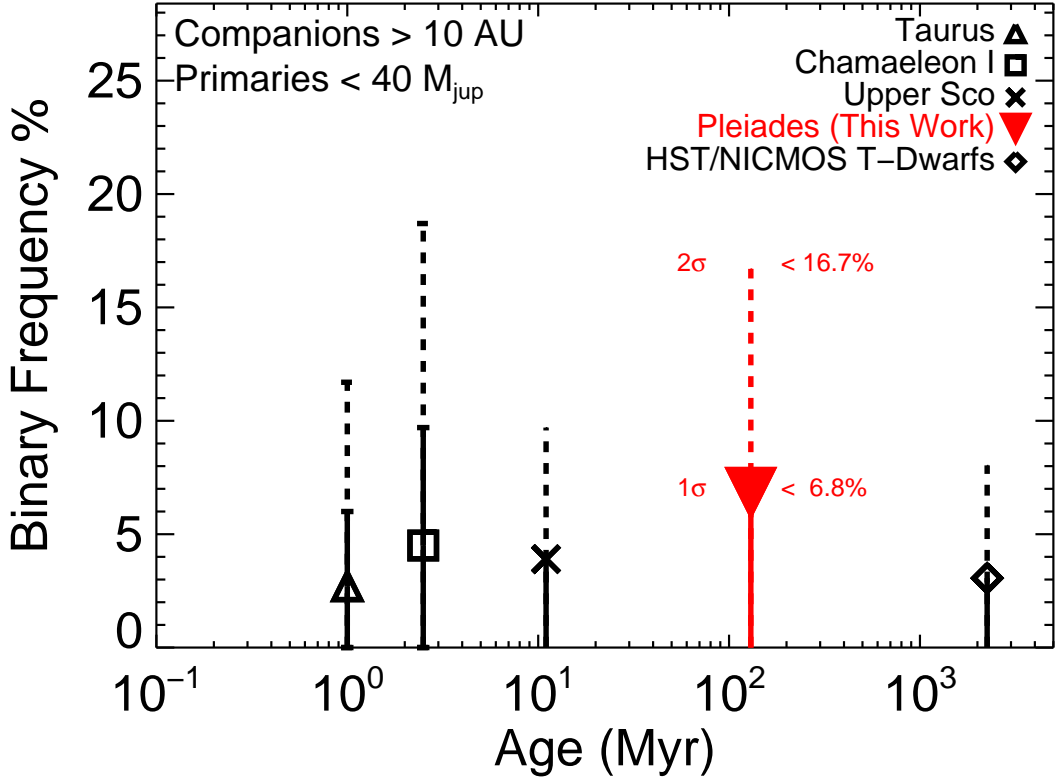}%{binary_vs_age_mjuplt50.ps} %make_binaryfreqvs_age.pro 
\end{center}
\caption{\label{fig:binvsage}
The wide ($>$10 AU) brown dwarf binary frequency as function of the age
for young star forming regions, the intermediate age Pleiades, 
and the field (see \S\ref{sec:binvsage}). All populations are shown for a 
common mass range ($25-40~\mjup$). Low mass brown dwarf binaries 
may very well be infrequent across a wide range of ages. }
\end{figure}

It is possible that 
the different cluster stellar densities in which brown dwarfs form could affect 
the binary frequency, hindering a direct comparison between field and young brown binary frequencies as done here.  
However, \cite{King12} find that the binary frequency for stars with masses of $0.1-0.3~\Msol$
did not vary measurably over nearly $20\times$ in density for five young regions (Taurus, Chamaeleon I, Ophiucus, IC 348, 
and the Orion Nebula Cluster). Figure \ref{fig:binvsage} and 
Table~\ref{table:binvsage} summarizes these comparisons 
of the binary frequency at different ages. In contrast to our estimate of the binary frequency in \S\ref{sec:binfrac}, 
here we used only the methods 
of \cite{Burgasser03} for computing the binary frequency
of these different clusters and the field in order to keep the statistical 
analysis the same. 

For constraining our binary frequency of Pleiades at 
wider separations $a\gtrsim10$ AU, 4 brown dwarfs observed by the {\it HST}/WFPC2 and 
{\it HST}/ACS surveys of \cite{Martin03} and \cite{Bouy06}
 were combined with our own observations for a larger sample size of $\nobjpleiades$ objects. 
These 4 brown dwarfs match our $K\gtrsim16.0$ mag cutoff
and conservative Pleiades cluster membership criteria, i.e.
that the target must have proper motion indicating cluster membership 
and a spectral type $\gtrsim$M9 (see \S\ref{sec:samp}). Brown dwarfs PLIZ 28 and PLIZ 2141 
were observed with {\it HST}/ACS by \cite{Bouy06} with detection limits that ruled out 
companions for mass ratios $q\gtrsim0.45$ at separations $a\gtrsim7-12$ AU. Brown dwarfs Roque 30 and Roque 33 
were observed with {\it HST}/WFPC2 by \cite{Martin03} and similarly they ruled 
out companions for mass ratios $q\gtrsim0.5$ and separations $a\gtrsim10$ AU. The {\it HST}/ACS and {\it HST}/WFPC2 
observations have comparable detection limits to our own detection limits of $q\gtrsim0.6$ at separations $a\gtrsim10$ AU. 
Thus, with a combined sample size of~\nobjpleiades~low mass Pleiades brown dwarfs and 
no binaries detected, we computed an upper limit on the binary 
frequency of $\binfreqpleiades\%$ ($1\sigma$) for mass ratios
$q\gtrsim0.6$ and separations $a\gtrsim10$ AU.

The sample of young brown dwarfs observed by {\it HST}/WFPC2 and AO surveys 
(see Table~\ref{table:binvsage}) compiled in \cite{Todorov14} and references 
therein includes all targets with spectral types $\gtrsim$M4. The detection limits for these surveys are generally sensitive to 
companions with separations $a\gtrsim10$ AU. 
In an attempt to constrain the masses of the primaries to $\lesssim$40$~\mjup$, 
we included only primaries in the \cite{Todorov14} sample with spectral types 
$\gtrsim$M7 (see Table~\ref{table:binvsage}). Note that for young ($<$10 Myr) 
brown dwarfs mass estimates at young ages are still uncertain %LUHMAN 03??
and could have large uncertainties due the lack of a well measured 
$T_{\rm eff}$ scale for these stars and uncertain atmospheric 
and stellar evolution models. This spectral type cut off corresponds to a 
mass estimate of $\lesssim$40$~\mjup$ 
at ages~$\approx\tautaurus$~Myr and $\approx\taucham$~Myr for the Taurus and Chamaeleon I regions, respectively,
according to the \cite{Baraffe03} models. Over 
this range there are $\nbintaurus$ out of $\nobjtaurus$ binaries in Taurus and $\nbincham$ out of $\nobjcham$
binaries in Chamaeleon I, which corresponds to binary 
frequencies of $\binfreqtaurus$\% and $\binfreqcham$\% ($1\sigma$) respectively. 
We find our binary frequency upper limit of $\binfreqpleiades\%$ 
is in agreement with binary frequencies for both Taurus and Chamaeleon I.
One caveat is we included candidate companions in 
Taurus 2MASS J04414489+2301513 and 2MASS J04221332+1934392  from \cite{Todorov14} in the 
binary frequency computed here. If those objects are not binaries, the binary frequency of 
Taurus would be even lower ($\binfreqtauruslower\%$), still in
agreement with our own binary frequency limit.  

\cite{Kraus12} and \cite{Biller11}  observed 10 and 18 members of Upper Sco with spectral types $\gtrsim$M7 respectively
and were sensitive to companions with separations $\gtrsim$$10$ AU. Given an age of 11\,Myr for Upper Sco \citep{Pecaut12} 
and the spectral type--$T_{\rm eff}$\ relation of \cite{Pecaut13}, 
$\gtrsim$M7 spectral types correspond to $\lesssim$2650\,K 
and thereby masses of $\lesssim$40$~\mjup$. This is comparable to our own mass range of~\massrange. 
Both previous surveys have detection limits $q\gtrsim0.8$ at separations $a\gtrsim10$ AU with no binaries detected. 
Using this combined sample, we estimated a binary frequency of $\binfrequppersco\%$ for Upper Sco, 
which is consistent to our own binary frequency 
upper limit of $\binfreqpleiades\%$ for the Pleiades. 

\cite{Burgasser06} resolved $\nbinfieldtwoau$ T dwarf binaries with separations of $a=1.8-5.0$ AU 
out of $\nobjfieldtwoau$ stars observed with {\it HST}/NICMOS. They computed a
Malmquist bias-corrected binary frequency of $\binfreqfieldtwoau$\% for mass ratios $q\gtrsim0.6$ and 
separations $a\gtrsim2$ AU. However, to directly compare to our detection limits, 
we recomputed their Malmquist bias-corrected binary fraction and considered only the $\nbinfieldfourau$ T dwarf binaries which 
have projected separations of $\gtrsim$$10$ AU, which gives a binary frequency of $\binfreqfieldtenau\%$ for 0 binaries detected out of $\nobjfieldtenau$ objects observed.  

\cite{Bate12} performed hydrodynamic simulations of star formation that 
produced 27 objects with masses $<$70~$\mjup$, with none ending up as binaries. \cite{Bate12} quoted a binary frequency of $0.0\pm5\%$ 
for the mass range of $30-70$~$\mjup$ and a binary frequency of $<$7$\%$ for the mass range $10-30$~$\mjup$. These predictions are
in good agreement with our observed binary frequency constraint of $\binfreqpleiades\%$ for separations 
$\gtrsim$$10$ AU. 

\section{Summary \label{sec:summ}} 

The measurement of the brown dwarf binary frequency at different ages 
is fundamental tool for testing theory, given that the
statistical properties of binaries probe formation scenarios in the very low-mass regime. 
In this work, we tripled the number low-mass Pleiades brown dwarfs searched for companions, 
surveying a sample of 11 previously unobserved L dwarfs in the Pleiades, predecessors to T dwarfs 
in the field, using {\it HST}/WFC3. 
We have constrained the binary frequency in Pleiades for the lowest known mass (\massrange) and latest 
known type ($\gtrsim$M9) brown dwarfs 
to $\binfreqjeff\%$ at $1\sigma$ ($\binfreqjefftwosig\%$ at $2\sigma$) confidence for companions as close as $\approx4$ AU, 
finding no binaries.  Our survey is the first to probe down to separations of $4$ AU at such young ages. 

Furthermore, we find our binary frequency constraints
are in good agreement with observed binary frequencies of young star forming regions 
Taurus ($\binfreqtaurus$\%), 
Chamaeleon I ($\binfreqcham$\%), and 
Upper Sco ($\binfrequppersco$\%) for objects with similar primary masses of $<$40$~$$\mjup$, at $1\sigma$ with 
projected separations $>$10 AU.  
Overall, our observations of the Pleiades support
the evidence that T dwarf binaries are likely uncommon, and consistent with having 
the same frequency at both young ($1-10$ Myr), intermediate ($\approx$$120$ Myr) and old ($\gtrsim$$1$ Gyr) ages.

\begin{deluxetable}{lccccccc}
\tabletypesize{\small}
\tablecolumns{8}
\tablewidth{0pt} 
\tablecaption{\label{table:samp} Pleiades Sample}
\tablehead{ 
\colhead{Name$^{a}$} & \colhead{R. A.} & \colhead{Decl.}	 &\colhead{Mass$^{b}$} & \colhead{K} & \colhead{SpT} & \colhead{SpT} & \colhead{P.M.} \\
\colhead{} & \colhead{J2000.0} & \colhead{J2000.0} &\colhead{$\mjup$} & \colhead{mag}  & \colhead{} & \colhead{Ref} & \colhead{Ref}
}
\startdata
BRB 17	 				& 03 54 07.98  		&	 +23 54 27.9 	 & 43 & $16.03\pm0.03$	 &  L0 &    1    &  2     \\      
NPNPL 2 					& 03 46 34.26 		&	 +23 50 03.7 	 & 41 & $16.09\pm0.03$	 &	 & 	&  	  	3	\\     	      	      
PLIZ 31 					&  03 51 47.65 		&	 +24 39 59.2 	 & 40 & $16.09\pm0.03$	 &	 &    	     &       	3,4          \\
BRB 21  					&	03 54 10.27 	&	 +23 41 40.2 	 & 31 & $16.39\pm0.04$	 &	   L3     & 1  	     &   	  	2  	\\	
PLIZ 35 					&      03 52 39.16 	&	 +24 46 29.5 	 & 31 & $16.51\pm0.04$	&	L2     & 1      		& 	  	2  	\\		
BRB 23  					& 03 50 39.54 		& 	+25 02 54.7 	 & 30 & $16.56\pm0.04$	 & 	L3.5   & 1     		&	 	2 	\\		
PLIZ 161 					& 03 51 29.47 		&	 +24 00 37.3 	 & 28 & $16.70\pm0.05$ 	  &    &             &	       3 	\\	
UGCS J0348+2550$^{e}$ 			&  03 48 15.63 		&	 +25 50 08.9 	 & 28 & $16.73\pm0.05$ 	  &  L3$\pm$1	 &     8    	 &    	3,7     	\\	
BRB 28 					&  03 52 54.90 		&	 +24 37 18.2 	& 26 & $16.92\pm0.06$	  &     &             & 	  	2  	\\	
PLIZ 1262 	 			&  03 44 27.27 		&	 +25 44 42.0 	 & 26 & $16.95\pm0.07$	  &	  &            	&     	2,4      \\ 
BRB 29 					&    03 54 01.43  	&	 +23 49 57.7 	 & 25 & $17.00\pm0.07$	  &      L4.5 &     1     		& 	  	2 \\			 
Roque 33$^{c}$ 			& 	03 48 49.03 	&	+24 20 25.4 	  & 41 & $16.06\pm0.03$ 	  &	 M9.5 & 6     	&      	  5	  	\\
Roque 30$^{c}$ 			&	03 50 16.09 	&	+24 08 34.7	   & 40 & $16.08\pm0.03$   &	  & 		 &       	  3\\	      	
PLIZ 28$^{d}$ 				&     03 54 14.03 	& 	 +23 17 51.4      & 35 & $16.14\pm0.03$ 	 & L0.0 & 1 	   		 &       	    2 	  \\
PLIZ 2141$^{d}$ 			&	03 44 31.29 	& 	+25 35 14.4      & 28 & $16.69\pm0.04$      &      &                 &               2   \\     
\enddata
\tablecomments{
$^{a}$ To search these targets by name in Simbad, add the string ``Cl* Melotte 22" \\
$^{b}$ Masses are estimated from \cite{Baraffe03} \\
$^{c}$ Observed with {\it HST}/WFPC2 \cite{Martin03} \\  
$^{d}$ Observed with {\it HST}/ACS \cite{Bouy06} \\ 
$^{e}$ UGCS J$034815.64+255008.9$ \\
{\bf References.} (1) \cite{Bihain10}; (2) \cite{Bihain06}; (3) \cite{Lodieu12}; (4) \cite{Casewell07}; (5) \cite{Stauffer07}; (6) \cite{Martin00}
(7) \cite{Zap14a}; (8) \cite{Zap14b}
}
\end{deluxetable}

\begin{deluxetable}{lll}
\tablecolumns{5}
\tablewidth{0pt} 
\tablecaption{\label{table:phot} {\it HST}/WFC3 Photometry}
\tablehead{ 
\colhead{Our Targets} & \colhead{F814W} & \colhead{F850LP} \\
\colhead{} & \colhead{(mag)}  & \colhead{(mag)} 
}
\startdata 
BRB 17  & $20.419\pm0.007$ & $19.415\pm0.011$  \\
NPNPL 2  &$20.685\pm0.006$& $19.448\pm0.011$ \\
PLIZ 31  &$20.701\pm0.006$& $19.524\pm0.013$ \\
BRB 21  &$21.344\pm0.010 $& $20.204\pm0.023$ \\
PLIZ 35 &$21.315\pm0.010 $&$20.096\pm0.021$ \\
BRB 23 &$21.604\pm0.012$& $20.431\pm0.029$\\
PLIZ 161  &$21.804\pm0.014$& $20.678\pm0.034$ \\ 
UGCS J0348+2550 &$21.866\pm0.015 $& $20.706\pm0.035$ \\
BRB 28   &$22.177\pm0.019$& $20.860\pm0.040$ \\
PLIZ 1262    &$22.211\pm0.020$&$21.086\pm0.049$ \\
BRB 29   &$22.231\pm0.021$& $21.042\pm0.048$ \\

\enddata
%\tablecomments{
%}
\end{deluxetable}

%\begin{deluxetable}{cccccc}
%\tablecolumns{4}
%\tablewidth{0pt} 
%\tablecaption{\label{table:sim} Binary Simulations}
%\tablehead{ 
%\colhead{Simulation} & \colhead{Filter} & \colhead{S/N} & \colhead{Possible} & \colhead{Number of}  \\ 
%\colhead{} & \colhead{} & \colhead{} & \colhead{Pairings} & \colhead{Artificial Binaries} }
%\startdata 
%Median S/N & F814W & \fwmedsn & \npfwmed & \nbinary  \\ 
%Min S/N & F814W & \fwminsn & \npfwmin & \nbinary   \\ 
%Median S/N & F850LP & \flpmedsn & \npflpmed & \nbinary   \\ 
%Min S/N & F850LP & \flpminsn & \npflpmin & \nbinary  \\
%\enddata
%\end{deluxetable}

\begin{deluxetable}{ccccc}
\tablecolumns{4}
\tablewidth{0pt} 
\tablecaption{\label{table:sim} Binary Simulations}
\tablehead{ 
\colhead{Simulation} & \colhead{Filter} & \colhead{S/N} & \colhead{Number of}  \\ 
\colhead{} & \colhead{} & \colhead{}  & \colhead{Artificial Binaries} }
\startdata 
Median S/N & F814W & \fwmedsn  & \nbinary  \\ 
Min S/N & F814W & \fwminsn & \nbinary   \\ 
Median S/N & F850LP & \flpmedsn & \nbinary   \\ 
Min S/N & F850LP & \flpminsn  & \nbinary  \\
\enddata
\end{deluxetable}

\begin{deluxetable}{ccccccccccc} %makecc_table.pro
\tablecolumns{11}
\tablewidth{0pt}
\tablecaption{\label{table:detectlim} Detection Limits}
\tablehead{
\colhead{Target} & \colhead{$0\farcs025$} &\colhead{$0\farcs032$} &\colhead{$0\farcs040$} &\colhead{$0\farcs050$} &\colhead{$0\farcs063$} &\colhead{$0\farcs080$} &\colhead{$0\farcs100$} &\colhead{$0\farcs126$} &\colhead{$0\farcs159$} &\colhead{$0\farcs180$}
}
\startdata
\multicolumn{10}{c}{{\bf F814W (mag)}} \\ \hline\hline
            BRB17 &      20.66 &      21.26 &      21.88 &      22.20 &      22.48 &      22.75 &      22.77 &      23.05 &      23.69 &      23.06 \\
           NPNPL2 &      20.93 &      21.52 &      22.15 &      22.46 &      22.75 &      23.02 &      23.04 &      23.31 &      23.95 &      23.33 \\
           PLIZ31 &      20.94 &      21.54 &      22.17 &      22.48 &      22.76 &      23.03 &      23.05 &      23.33 &      23.97 &      23.35 \\
            BRB21 &      21.59 &      22.18 &      22.81 &      23.12 &      23.41 &      23.67 &      23.70 &      23.97 &      24.61 &      23.99 \\
           PLIZ35 &      21.56 &      22.15 &      22.78 &      23.09 &      23.38 &      23.65 &      23.67 &      23.94 &      24.58 &      23.96 \\
            BRB23 &      21.85 &      22.44 &      23.07 &      23.38 &      23.67 &      23.93 &      23.96 &      24.23 &      24.87 &      24.25 \\
          PLIZ161 &            &      22.64 &      23.27 &      23.26 &      23.54 &      24.13 &      23.86 &      24.43 &      24.72 &      24.45 \\
   UGCSJ0348+2550 &            &      22.70 &      23.33 &      23.32 &      23.61 &      23.93 &      23.92 &      24.49 &      24.78 &      24.51 \\
            BRB28 &            &      22.70 &      23.30 &      23.63 &      23.92 &      24.25 &      24.23 &      24.53 &      24.76 &      24.55 \\
         PLIZ1262 &            &      22.73 &      23.34 &      23.66 &      23.66 &      24.28 &      24.27 &      24.57 &      24.80 &      24.58 \\
            BRB29 &            &      22.75 &      23.36 &      23.68 &      23.68 &      24.30 &      24.29 &      24.59 &      24.82 &      25.44 \\
            \\
\multicolumn{10}{c}{{\bf F850LP (mag)}} \\ \hline\hline
            BRB17 &            &            &      20.26 &      20.55 &      20.57 &      20.90 &      21.49 &      21.14 &      21.77 &      21.76 \\
           NPNPL2 &            &            &      20.29 &      20.59 &      20.60 &      20.93 &      21.52 &      21.17 &      21.81 &      21.80 \\
           PLIZ31 &            &            &      20.37 &      20.66 &      20.68 &      21.01 &      21.60 &      21.25 &      21.88 &      21.87 \\
            BRB21 &            &            &      21.05 &      21.34 &      21.36 &      21.69 &      22.28 &      21.93 &      22.56 &      22.55 \\
           PLIZ35 &            &            &      20.94 &      21.23 &      21.25 &      21.58 &      22.17 &      21.82 &      22.46 &      22.44 \\
            BRB23 &            &            &      21.28 &      21.57 &      21.59 &      21.91 &      22.51 &      22.16 &      22.79 &      22.78 \\
          PLIZ161 &            &            &      20.73 &      20.92 &      21.83 &      22.16 &      22.12 &      22.40 &      22.44 &      22.16 \\
   UGCS J0348+2550 &            &            &      20.76 &      20.95 &      21.86 &      22.19 &      22.15 &      22.43 &      22.46 &      22.19 \\
            BRB28 &            &            &            &      20.91 &      21.71 &      22.02 &      22.31 &      22.30 &      22.30 &      22.34 \\
         PLIZ1262 &            &            &            &      21.14 &      21.94 &      22.24 &      22.53 &      22.53 &      22.53 &      22.57 \\
            BRB29 &            &            &            &      21.10 &      21.90 &      21.88 &      22.49 &      22.48 &      22.48 &      22.53 \\
\enddata
%\tablecomments{Detection limits here are target mag photometry (\S\ref{sec:hstwfc3imaging}) added to out contrast curve $\Delta\rm m$ (\S\ref{sec:cc}). 
%}
\end{deluxetable}

%\begin{deluxetable}{l l l l l l |}
\begin{deluxetable}{cccccccccc}
\tablecolumns{10}
\tablewidth{0pt} 
\tablecaption{\label{table:binvsage} Binary Frequency vs. Age for Wide ($>$10) AU Companions}

\tablehead{
\colhead{Region} & \colhead{Age} & \colhead{Age}            & \colhead{Sample} & \colhead{$N_{\rm obj}$}   & \colhead{$N_{\rm bin}$} & \colhead{Bin Freq}    & \colhead{$q$} \\ 
\colhead{}            &  \colhead{}       & \colhead{Ref}       & \colhead{Ref}        & \colhead{}                         & \colhead{}                       & \colhead{\%}               & \colhead{}
}
\startdata
Taurus & \tautaurus~Myr  					& 15         			& 1,2,3,4,5 & \nobjtaurus & \nbintaurus & $\binfreqtaurus$ 		    & \qrngtaurus \\
Chameleon I & \taucham~Myr 				& 16 					& 4,5,6,7,8,9,10 & \nobjcham & \nbincham & $\binfreqcham$ 		    & \qrngcham \\
Upper Sco & \tauuppersco~Myr 			& 17 		   			& 2,11 &\nobjuppersco & \nbinuppersco & $\binfrequppersco$ 		     & \qrnguppersco \\
This work + lit & \taupleiades~Myr 	                 & 18 						& 12,13 &\nobjpleiades & \nbinpleiades & $\binfreqpleiades$ 		      & \qrngpleiades \\
Field & \taufield~Gyr 					& 19                                & 14 & \nobjfieldtenau & \nbinfieldtenau & $\binfreqfieldtenau$  & \qrngfield \\
\enddata
\tablecomments{Faint companions to brown dwarfs with separations and mass ratios 
greater than given in table are ruled out by the given detection limits for primaries with masses $<$40$\mjup$ and separations $>$10 AU.  \\
{\bf References.} 
(1) \cite{Todorov14}; (2) \cite{Kraus12}; (3) \cite{Kraus06}(4) \cite{Konopacky07} (5) \cite{Todorov10}; %TAURUS
(6) \cite{Luhman04}; (7) \cite{Laf08}; (8) \cite{Ahmic07}; (9) \cite{Luhman07}; (10) \cite{Neuhauser02}; %CHAMAELEONI
(11) \cite{Biller11};  %UPPERSCO
(12) \cite{Martin03}; (13) \cite{Bouy06}; %PLEIADES
(14) \cite{Burgasser06}; %THE FIELD
(15) \cite{Luhman07}; %TAURUS AGE
(16) \cite{Luhman10}; %CHAMAELEON I AGE
(17) \cite{Pecaut12}; %UPPER SCO AGE
(18) \cite{Barrado04}; %PLEIADES AGE 
(19) Assumed age for field T dwarfs by \cite{Burgasser06} from \cite{ReidHawley00}. %FIELD AGE
}
\end{deluxetable}

\begin{deluxetable}{cc}
\tablecolumns{2}
\tablewidth{0pt} 
\tablecaption{\label{table:compmapprob} Companion Detectability}
\tablehead{
\colhead{Name} & \colhead{Detectability} \\ %& \colhead{uniform $a$}% \\
\colhead{} & \colhead{log normal $a$} \\%& \colhead{$P(q)\propto q^{4.9}$} \\
}
\startdata
BRB17 & $70.8$\% \\%& $74.9$\% \\
BRB21 & $68.7$\% \\%& $73.8$\% \\
BRB23 & $69.9$\% \\%& $74.4$\% \\
BRB28 & $67.2$\% \\%& $71.8$\% \\
BRB29 & $66.7$\% \\%& $72.0$\% \\
NPNPL2 & $71.4$\% \\%& $75.6$\% \\
PLIZ1262 & $66.9$\% \\%& $71.7$\% \\
PLIZ161 & $68.8$\% \\%& $73.3$\% \\
PLIZ31 & $71.3$\% \\%& $75.8$\% \\
PLIZ35 & $68.6$\% \\%& $73.7$\% \\
UGCS J0348+2550 & $68.8$\% \\%& $73.3$\% \\
Total Expected Binaries & $  7.6$ \\%& $  8.1$ \\
Binary Frequency$^{a}$ & $\binfreqjeff$\% \\%& \\
\enddata
\tablecomments{$^{a}$ Binary frequency with $1\sigma$ using the Jeffrey interval recommended for low $n$ by \cite{Brown01}.}
\end{deluxetable}

\end{document}